\documentclass[aps,prl,twocolumn,floatfix,superscriptaddress]{revtex4}
\usepackage{amsmath}
\usepackage{graphicx}
\usepackage{bm}
\begin{document}
\title{What is the most competitive sport?}
\author{E.~Ben-Naim}
\email{ebn@lanl.gov}
\affiliation{Theoretical Division and Center for Nonlinear Studies,
Los Alamos National Laboratory, Los Alamos, New Mexico 87545}
\author{F.~Vazquez}
\email{fvazquez@buphy.bu.edu}
\affiliation{Theoretical Division and Center for Nonlinear Studies,
Los Alamos National Laboratory, Los Alamos, New Mexico 87545}
\affiliation{Department of Physics, Boston University, Boston,
Massachusetts 02215}
\author{S.~Redner}
\email{redner@bu.edu}
\affiliation{Theoretical Division and Center for Nonlinear Studies,
Los Alamos National Laboratory, Los Alamos, New Mexico 87545}
\affiliation{Department of Physics, Boston University, Boston,
Massachusetts 02215}
\begin{abstract}
We present an extensive statistical analysis of the results of all
sports competitions in five major sports leagues in England and the
United States. We characterize the parity among teams by the variance
in the winning fraction from season-end standings data and quantify
the predictability of games by the frequency of upsets from game
results data. We introduce a mathematical model in which the underdog
team wins with a fixed upset probability. This model quantitatively
relates the parity among teams with the predictability of the games,
and it can be used to estimate the upset frequency from standings
data. We propose the likelihood of upsets as a measure of
competitiveness.
\end{abstract}

\maketitle

What is the most competitive team sport?  We answer this question via
a statistical survey of game results \cite{s,gts,abc,pn}. We relate
{\it parity} with {\it predictability}, and propose the likelihood of
upsets as a measure of competitiveness.

We studied the results of all regular season competitions in 5 major
professional sports leagues in England and the United States (table
I): the premier soccer league of the English Football Association
(FA), Major League Baseball (MLB), the National Hockey League (NHL),
the National Basketball Association (NBA), and the National Football
League (NFL). NFL data includes the short-lived AFL. We considered
only complete seasons, with more than 300,000 games in over a century
\cite{data}.

The winning fraction, the ratio of wins to total games, quantifies
team strength.  Thus the distribution of winning fraction measures the
parity between teams in a league.  We computed $F(x)$, the fraction of
teams with a winning fraction of $x$ or lower at the end of the
season, as well as $\sigma=\sqrt{\langle x^2\rangle -\langle
x\rangle^2}$, the standard deviation in winning fraction. Here
$\langle\cdot\rangle$ denotes the average over all teams and all years
using season-end standings. For example, in baseball where the winning
fraction $x$ typically falls between $0.400$ and $0.600$, the variance
is $\sigma=0.084$. As shown in figures 1 and 2a, the winning fraction
distribution clearly distinguishes the five leagues. It is narrowest
for baseball and widest for football.

\begin{table}[b]
\begin{tabular}{|l|c|c|c|c|c|c|}
\hline
league&years&games&$\langle{\rm games}\rangle$&$\sigma$&${\bm q}$&$q_{\rm model}$\\
\hline
FA&1888-2005&43350&39.7&0.102&{\bf 0.452}&0.459\\
MLB&1901-2005&163720&155.5&0.084&{\bf 0.441}&0.413\\
NHL&1917-2004&39563&70.8&0.120&{\bf 0.414}&0.383\\
NBA&1946-2005&43254&79.1&0.150&{\bf 0.365}&0.316\\
NFL&1922-2004&11770&14.0&0.210&{\bf 0.364}&0.309\\
\hline
\end{tabular}
\caption{Summary of the sports statistics data. Listed are the time
  periods, total number of games, average number of games played by a
  team in a season ($\langle{\rm games}\rangle$), variance in the
  win-percentage distribution ($\sigma$), measured frequency of upsets
  (${\bm q}$), and upset probability obtained using the theoretical
  model ($q_{\rm model}$). The fraction of ties in soccer, hockey, and
  football is $0.246$, $0.144$, and $0.016$, respectively.}
\label{table}
\end{table}

Do these results imply that MLB games are the most competitive and NFL
games the least? Not necessarily! The length of the season is a
significant factor in the variability in the winning fraction.  In a
scenario where the outcome of a game is completely random, the total
number of wins performs a simple random walk, and the standard
deviation $\sigma$ is inversely proportional to the square root of the
number of games played. Generally, the shorter the season, the larger
$\sigma$. Thus, the small number of games is partially responsible for
the large variability observed in the NFL.

\begin{figure}[t]
\vspace*{0.01cm}
\includegraphics*[width=0.37\textwidth]{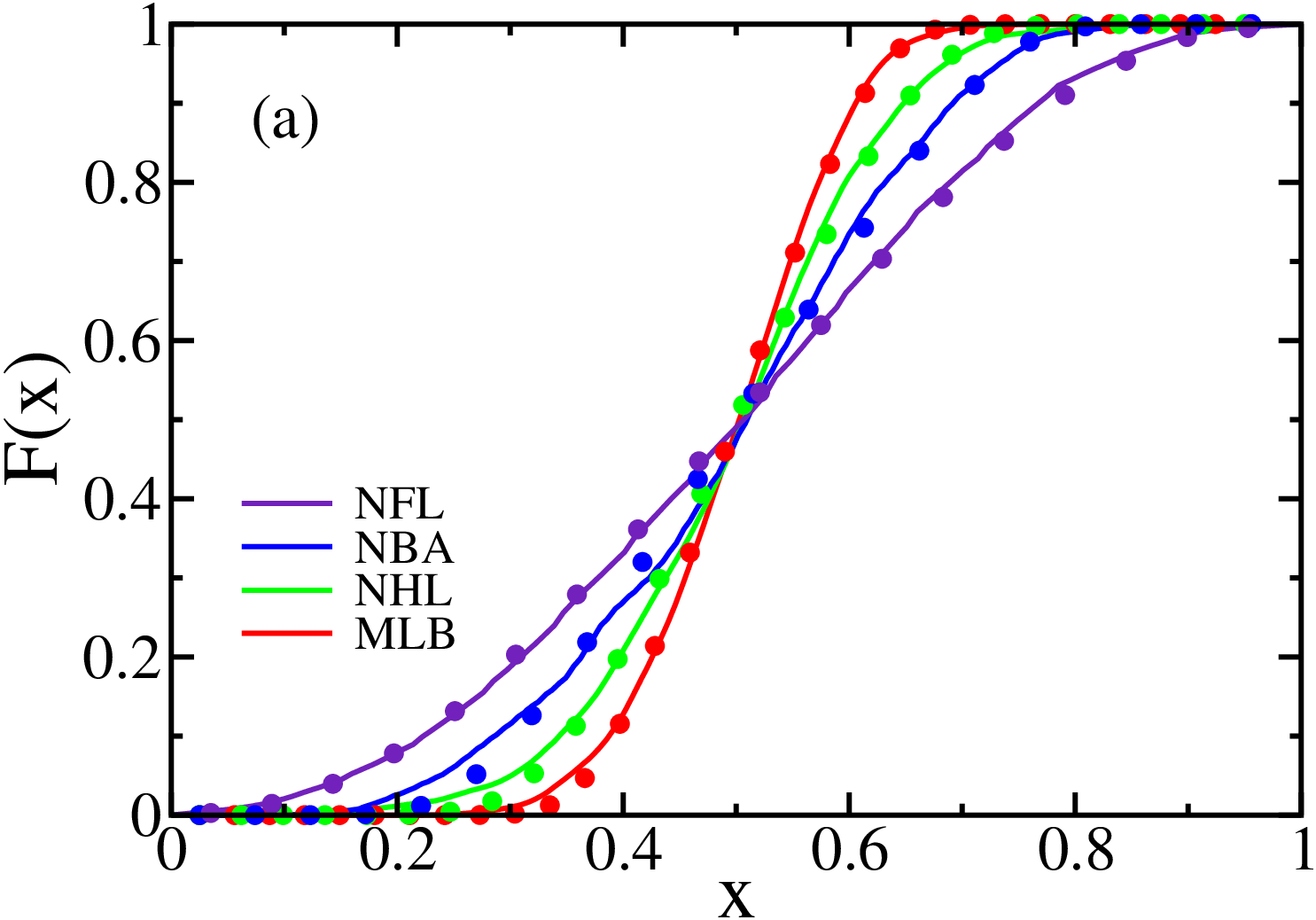}
\caption{Winning fraction distribution (curves) and the best-fit
distributions from simulations of our model (circles). For clarity,
FA, that lies between MLB and NHL, is not displayed.}
\end{figure}

To account for the varying season length and reveal the true nature of
the sport, we set up mock sports leagues where teams, paired at
random, play a fixed number of games. In this simulation model, the
team with the better record is considered as the favorite and the team
with the worse record is considered as the underdog. The outcome of a
game depends on the relative team strengths: with the ``upset
probability'' $q<1/2$, the underdog wins, but otherwise, the favorite
wins.  Our analysis of the nonlinear master equations that describe
the evolution of the distribution of team win/loss records shows that
$\sigma$ decreases both as the season length increases and as games
become more competitive, i.e., as $q$ increases \cite{bvr}. In a
hypothetical season with an infinite number of games, the winning
fraction distribution is uniform in the range $q<x<1-q$ and as a
result, $\sigma=(1/2-q)/\sqrt{3}$.

We run Monte Carlo simulations of these artificial sports leagues,
with sport-specific number of games and a range of $q$ values.  We
then determine the value of $q$ that gives the best match between the
distribution $F(x)$ from the simulations to the actual sports
statistics (figure 1).  Generally, we find good agreement between the
simulations results and the data for reasonable $q$ values.

\begin{figure}[t]
\vspace*{0.01cm}
\includegraphics*[width=0.37\textwidth]{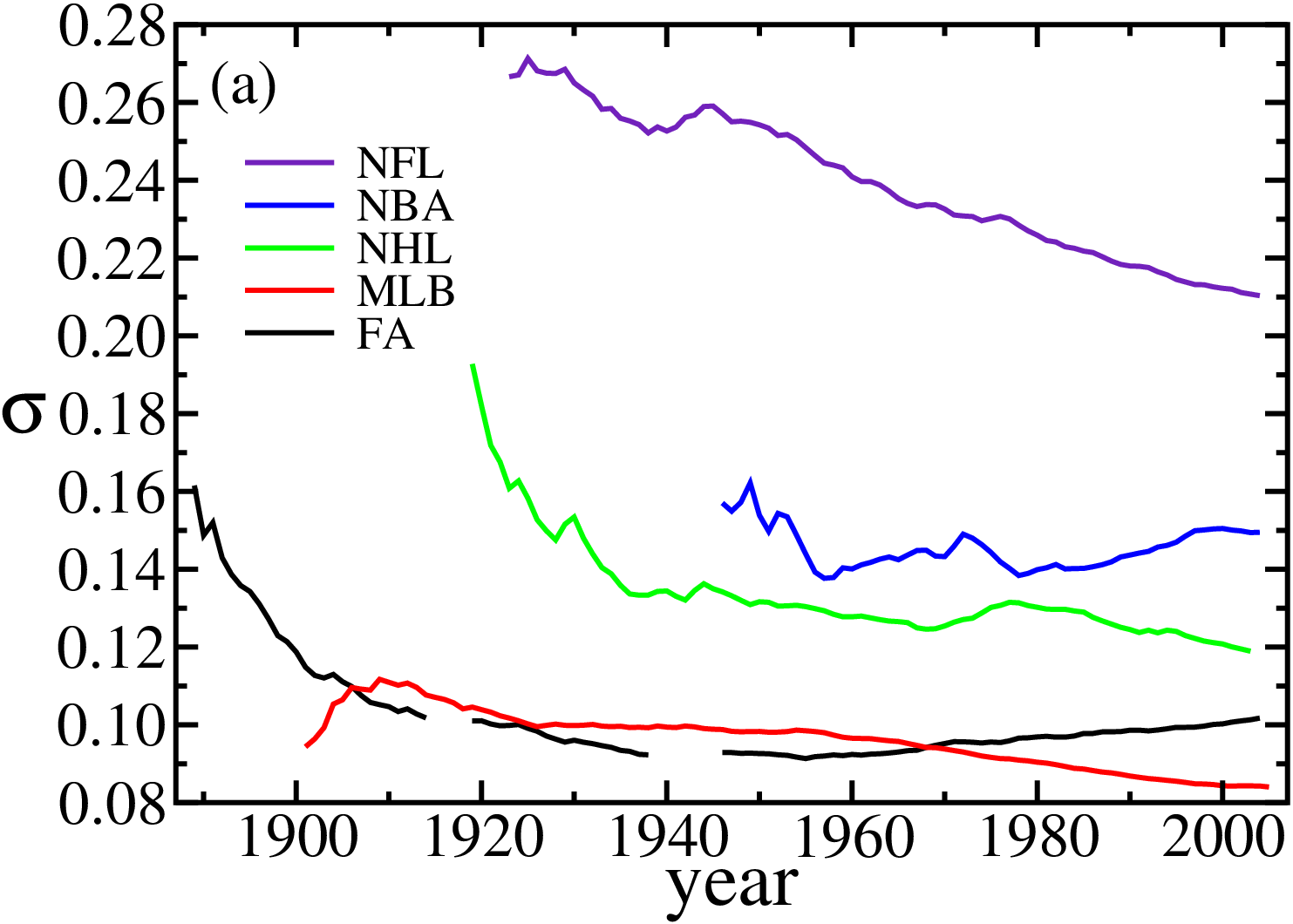}
\includegraphics*[width=0.37\textwidth]{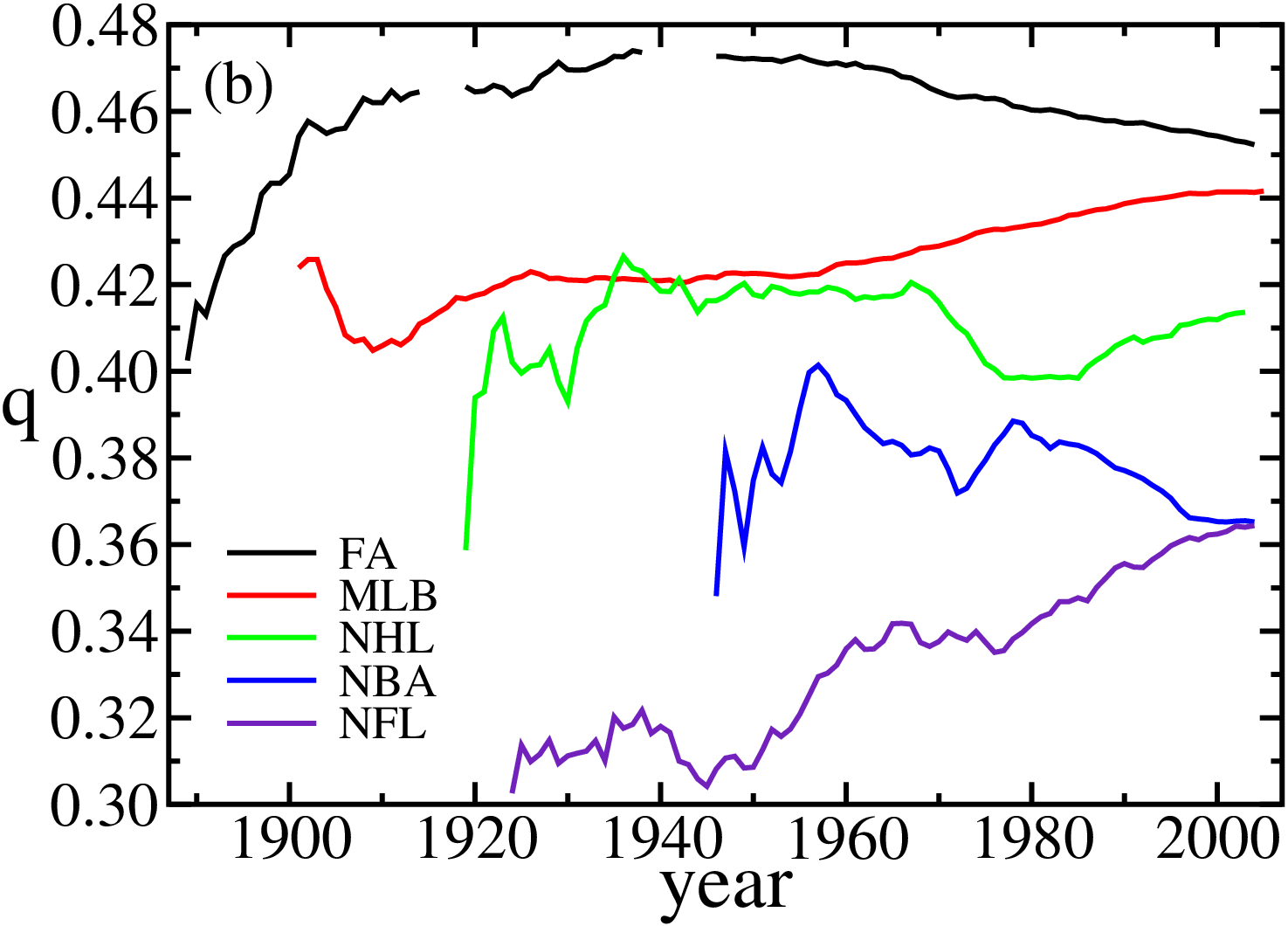}
\caption{(a) The cumulative variance in the winning fraction distribution (for
all seasons up to a given year) versus time.  (b) The cumulative frequency of
upsets $q$, measured directly from game results, versus time.}
\end{figure}

To characterize the predictability of games, we followed the
chronologically-ordered results of all games and reconstructed the
league standings at any given day. We then measured the upset
frequency $q$ by counting the fraction of times that the team with the
worse record on the game date actually won (table I). Games between
teams with no record (start of a season) or teams with equal records
were disregarded. Game location was ignored and so was the margin of
victory. In soccer, hockey, and football, ties were counted as 1/2 of
a victory for both teams.  We verified that handling ties this way did
not significantly affect the results: the upset probability changes by
at most $0.02$ (and typically, much less) if ties are ignored.

We find that soccer and baseball are the most competitive sports with
$q=0.452$ and $q=0.441$, respectively, while basketball and football,
with nearly identical $q=0.365$ and $q=0.364$, are the least.  There
is also good agreement between the upset probability $q_{\rm model}$,
obtained by fitting the winning fraction distribution from numerical
simulations of our model to the data as in figure 1, and the measured
upset frequency (table I). Consistent with our theory, the variance
$\sigma$ mirrors the bias, $1/2-q$ (figures 2a and 2b).  Tracking the
evolution of either $q$ or $\sigma$ leads to the same conclusion: NFL
and MLB games are becoming more competitive, while over the past 60
years, FA displays an opposite trend.

In summary, we propose a single quantity, $q$, the frequency of
upsets, as an index for quantifying the predictability, and hence the
competitiveness of sports games.  We demonstrated the utility of this
measure via a comparative analysis that shows that soccer and baseball
are the most competitive sports. Trends in this measure may reflect
the gradual evolution of the teams in response to competitive pressure
\cite{G}, as well as changes in game strategy or rules \cite{hs}.

Our model, in which the stronger team is favored to win a game
\cite{bvr}, enables us to take into account the varying season length
and this model directly relates parity, as measured by the variance
$\sigma$ with predictability, as measured by the upset likelihood
$q$. This connection has practical utility as it allows one to
conveniently estimate the likelihood of upsets from the more
easily-accessible standings data. In our theory, all teams are equal
at the start of the season, but by chance, some end up strong and some
weak. Our idealized model does not include the notion of innate team
strength; nevertheless, the spontaneous emergence of
disparate-strength teams provides the crucial mechanism needed for
quantitative modeling of the complex dynamics of sports competitions.

\smallskip
We thank Micha Ben-Naim for assistance in data collection and
acknowledge support from DOE (W-7405-ENG-36) and NSF (DMR0227670 \&
DMR0535503).

\end{document}